\documentclass[a4paper,fleqn,usenatbib]{mnras}
\usepackage[T1]{fontenc}
\usepackage{ae,aecompl}

\usepackage[dvipdfmx]{graphicx}	% Including figure files
\usepackage[dvipdfmx]{color}
\usepackage{amsmath}	% Advanced maths commands
\usepackage{amssymb}	% Extra maths symbols
\usepackage{lscape}

\title[Stellar Winds from Accreting Supermassive Stars]
{Do Stellar Winds Prevent the Formation of Supermassive Stars by Accretion?}

\author[D. Nakauchi et al.]{
Daisuke Nakauchi$^{1}$\thanks{E-mail: nakauchi@astr.tohoku.ac.jp},
Takashi Hosokawa$^{2, 3, 4}$, 
Kazuyuki Omukai$^{1, 4}$,
\newauthor \ Hideyuki Saio$^{1}$
and Ken'ichi Nomoto$^{5}$\thanks{Hamamatsu Professor}
\\
$^1$Astronomical Institute, Tohoku University, Aoba, Sendai 980-8578, Japan\\
$^2$Department of Physics, Kyoto University, Oiwake-cho, Kitashirakawa, Sakyo, Kyoto 606-8502, Japan\\
$^3$Department of Physics, The University of Tokyo, Bunkyo, Tokyo 113-0033, Japan\\
$^4$Kavli Institute for Theoretical Physics, Kohn Hall, University of California, Santa Barbara, CA 93106, USA\\
$^5$Kavli Institute for the Physics and Mathematics of the Universe (WPI), The University of Tokyo, Kashiwa, Chiba 277-8583, Japan\\
}

\date{Accepted XXX. Received YYY; in original form ZZZ}

\pubyear{2017}

\begin{document}
\label{firstpage}
\pagerange{\pageref{firstpage}--\pageref{lastpage}}
\maketitle

\begin{abstract}
Supermassive stars~(SMS; $\sim 10^5\ {\rm M}_\odot$) formed from metal-free gas in the early Universe attract attention as progenitors of supermassive black holes observed at high redshifts.
To form SMSs by accretion, central protostars must accrete at as high rates as $\sim 0.1\mbox{-}1\ {\rm M}_\odot\ {\rm yr}^{-1}$.
Such protostars have very extended structures with bloated envelopes, like super-giant stars, and 
are called super-giant protostars~(SGPSs).
Under the assumption of hydrostatic equilibrium, SGPSs have density inverted layers, where the luminosity becomes
locally super-Eddington, near the surface.
If the envelope matter is allowed to flow out, however, a stellar wind could be launched and
hinder the accretion growth of SGPSs before reaching the supermassive regime.
We examine whether radiation-driven winds are launched from SGPSs 
by constructing steady and spherically symmetric wind solutions. 
We find that the wind velocity does not reach the escape velocity in any case considered. 
This is because once the temperature falls below $\sim 10^4\ {\rm K}$, the opacity plummet drastically 
owing to the recombination of hydrogen and the acceleration ceases suddenly. 
This indicates that, in realistic non-steady cases, even if outflows are launched from the surface of 
SGPSs, they would fall back again. 
Such a ``wind'' does not result in net mass loss and does not prevent the growth of SGPSs.
In conclusion, SGPSs will grow to SMSs 
and eventually collapse to massive BHs of $\sim 10^5\ {\rm M}_\odot$, 
as long as the rapid accretion is maintained. 
\end{abstract}


%%%%%%%%%%%%%%%%%%%%%%%%%%%%%%%%%%%%%%%%%%%%%%%%%%

%%%%%%%%%%%%%%%%% BODY OF PAPER %%%%%%%%%%%%%%%%%%

\section{Introduction}
In the last decade, a number of luminous quasars~(QSOs) have been discovered at redshifts 
greater than 6~\citep{Fan2006,Mortlock2011,Venemans2013,Wu2015}, 
including the current record holder ULAS J1120+0641 at $z = 7.1$~\citep{Mortlock2011}.
This means that supermassive black holes~(SMBHs) of $\sim 10^9\mbox{-}10^{10}\ {\rm M}_\odot$ 
have already existed in less than a billion year after the Big Bang.
Such early formation poses a challenge to theories of the SMBH formation~\citep[e.g.,][]{Volonteri2010, Haiman2013}.

Although the first stars are considered theoretically to be typically massive with 
$\sim100\ {\rm M}_\odot$, and even can be as massive as 
$\sim1000\ {\rm M}_\odot$ in some circumstances, \citep[e.g.,][]{Hosokawa2011, Hirano2014}, it takes 0.84 and 0.73 Gyr 
for their remnant BHs of $100$ and $1000\ {\rm M}_\odot$, respectively, to reach the mass of 
the $z = 7.1$ SMBH, $2 \times 10^9\ {\rm M}_\odot$, via the Eddington-limited accretion. 
These growth time scales are still exceeding (for seed BHs of $100\ {\rm M}_\odot$) 
or only slightly below (for $1000\ {\rm M}_\odot$ seeds) the age of the Universe at that time, $0.77$ Gyr.
Even in the latter case, the BH is required to continuously accrete at the Eddington rate 
all the way to the SMBH, i.e., the 100 \% duty cycle in the six orders of magnitude in mass, 
which is quite improbable both from the observational and theoretical points of view.
From the high-$z$ QSO observations, the duty cycle is estimated as $\lesssim$ 60 \% 
at most at $z \geq 3.5$~\citep[e.g.,][]{Shen2007, Shankar2010}.
Theoretically, radiative feedback from the BH will make the growth time longer, 
so that the situation becomes even worse~\citep[e.g.,][]{Alvarez2009,Milos2009}, 
although the super-Eddington accretion, if it occurred, may help shorten the growth 
time enormously~\citep[e.g.,][]{Volonteri2005, Alexander2014}.

The so-called direct collapse scenario is an alternative pathway.
In this framework, we suppose that a supermassive star~(SMS) of $\sim 10^5\ {\rm M}_\odot$ 
forms from metal-free gas in the early Universe and collapses directly to a BH with almost the same mass by the post-Newtonian instability~\citep[e.g.,][]{Shapiro1983}.
With the more massive seeds, the growth time to $\sim 2 \times 10^9\ {\rm M}_\odot$ is 
reduced to $< 0.5\ {\rm Gyr}$, below the age of the Universe at $z = 7.1$ by some margin.

In a currently favored scenario~\citep{Bromm_Loeb2003}, the SMSs are supposed to be formed in atomic-cooling halos where the H$_2$ formation is prohibited either by photodissociation due to strong far-ultraviolet radiation~\citep{Omukai2001, Wolcott-Green2011b, Sugimura2014} or collisional dissociation by a high-density shock~\citep{Inayoshi2012}.
In such halos, a cloud collapses isothermally at $\sim$ 8000 K solely by the atomic cooling~\citep{Omukai2001}.
Without a major episode of cooling, the cloud collapses monolithically 
avoiding significant fragmentation until the formation of a protostar at its center~\citep{Inayoshi_Omukai2014, Becerra2015}.
The high temperature in the pre-stellar cloud results in the high accretion rate of 
$\dot{M}_{\rm acc} = 0.1\mbox{-}1\ {\rm M}_{\odot}\ {\rm yr}^{-1}$ onto the protostar 
according to the relation $\dot{M}_{\rm acc} \sim c_{\rm s}^3/G$~\citep[e.g.,][]{Shu1977}.
Note that even a tiny amount of metals induces significant fragmentation in the collapsing gas cloud, so that SMS formation can proceed only in the metal-free environment~\citep{Omukai2008}.

Such rapid accretion must be maintained until the central protostar grows to 
$\gtrsim 10^5\ {\rm M}_\odot$ by circumventing the possible obstacles.
For example, in the case of the formation of ordinary first stars, radiative feedback, including the photoevaporation of the accretion flows, 
plays an important role in terminating their accretion growth and setting the final mass 
at a few 10-100 ${\rm M}_{\odot}$~\citep[e.g.,][]{McKee2008, Hosokawa2011, Hosokawa2016, Susa2013}.
But, this is not the case for the SMS formation.
With the accretion rate exceeding a threshold value, 
$0.03\ {\rm M}_{\odot}\ {\rm yr}^{-1}$, the protostellar evolution changes completely~\citep{Hosokawa2012b, Hosokawa2013}.
Once the protostellar luminosity becomes close to the ``classical" Eddington luminosity, 
$L_{\rm Edd, es} = 4 \pi c G M_\ast / \kappa_{\rm es}$ where $\kappa_{\rm es}$ is the Thomson scattering opacity, 
at a few $10\ {\rm M}_{\odot}$, the stellar envelope swells greatly in radius reaching as large as 10-100 AU.
With the stellar effective temperature as low as $\sim 5000\ {\rm K}$, 
UV photons are hardly emitted and radiative feedback is too weak to halt the accretion.
Resembling the present-day red super-giant stars in appearance, the name ``super-giant protostars'' (SGPSs) 
is coined for the rapidly accreting stars with the bloated envelopes.
It is also known that the pulsational mass-loss rates from SGPSs are at most 
$\sim 10^{-3}\ {\rm M}_{\odot}\ {\rm yr}^{-1}$, two or three orders of magnitude 
lower than the accretion rate~\citep{Inayoshi2013}.
Thus the pulsation either would not prevent them growing supermassive.

A radiation driven stellar wind is another possible obstacle for the SGPS growth.
Similarly to the local Wolf-Rayet~(WR) stars, which exhibit the mass-loss at the rates of 
$\dot{M}_{\rm w} \sim 10^{-5}\mbox{-}10^{-4}\ {\rm M}_\odot\ {\rm yr}^{-1}$ in radiation-driven winds~\citep{Grafener2012}, 
the SGPSs have luminosities close to the classical Eddington value.
In addition, the SGPSs have a layer of density inversion, where the density increases outwardly, near the surface.
Although the radiative luminosity locally exceeds the Eddington value 
$L_{\rm Edd, local} = 4 \pi c G M_\ast / \kappa$, where $\kappa$ is the local 
opacity~\citep{Hosokawa2012b, Hosokawa2013}, the hydrostatic equilibrium is still
achieved as the layer is pushed down by the weight of the outer dense layers~\citep[e.g.,][]{Joss1973}.
If we omit the assumption of hydrostatic equilibrium and allow the matter 
to flow, however, we may find a wind solution blowing from the stellar surface~\citep{Ro2016}.
If such a stellar wind causes the significant mass loss, the stellar mass growth via accretion
may be stopped at some moment before the formation of a SMS.
To examine such a possibility, we here construct steady stellar wind solutions launched 
from the surface of SGPSs assuming the spherical symmetry.
We find that the radiation pressure force in fact allows 
the smooth acceleration from the subsonic to supersonic regime.
The wind velocity, however, does not reach the stellar escape velocity 
since the acceleration is suddenly over due to the opacity cutoff below $\sim 10^4$ K.
We thus conclude that the stellar wind either does not prevent the growth of a SGPS 
and it will eventually grow to a SMS as long as the rapid accretion is maintained.

The rest of this paper is organized as follows.
In Section \ref{sec:wind_formulation}, we describe the basic equations and the method to construct the stellar wind models.
In Section \ref{sec:wind_general}, we construct a series of wind solutions passing through the sonic point smoothly, 
without considering the connection to the stars at their bases and classify the solutions.
In Section \ref{sec:wind_match}, we present the wind solutions connected to the SGPSs and examine whether the wind mass-loss occurs 
from the SGPSs. 
Finally, Section \ref{sec:summary} is devoted to the summary and discussion.

\section{Formulation of Optically Thick Wind Solutions}\label{sec:wind_formulation}

\begin{figure}
\begin{center}
\includegraphics[scale = 0.3]{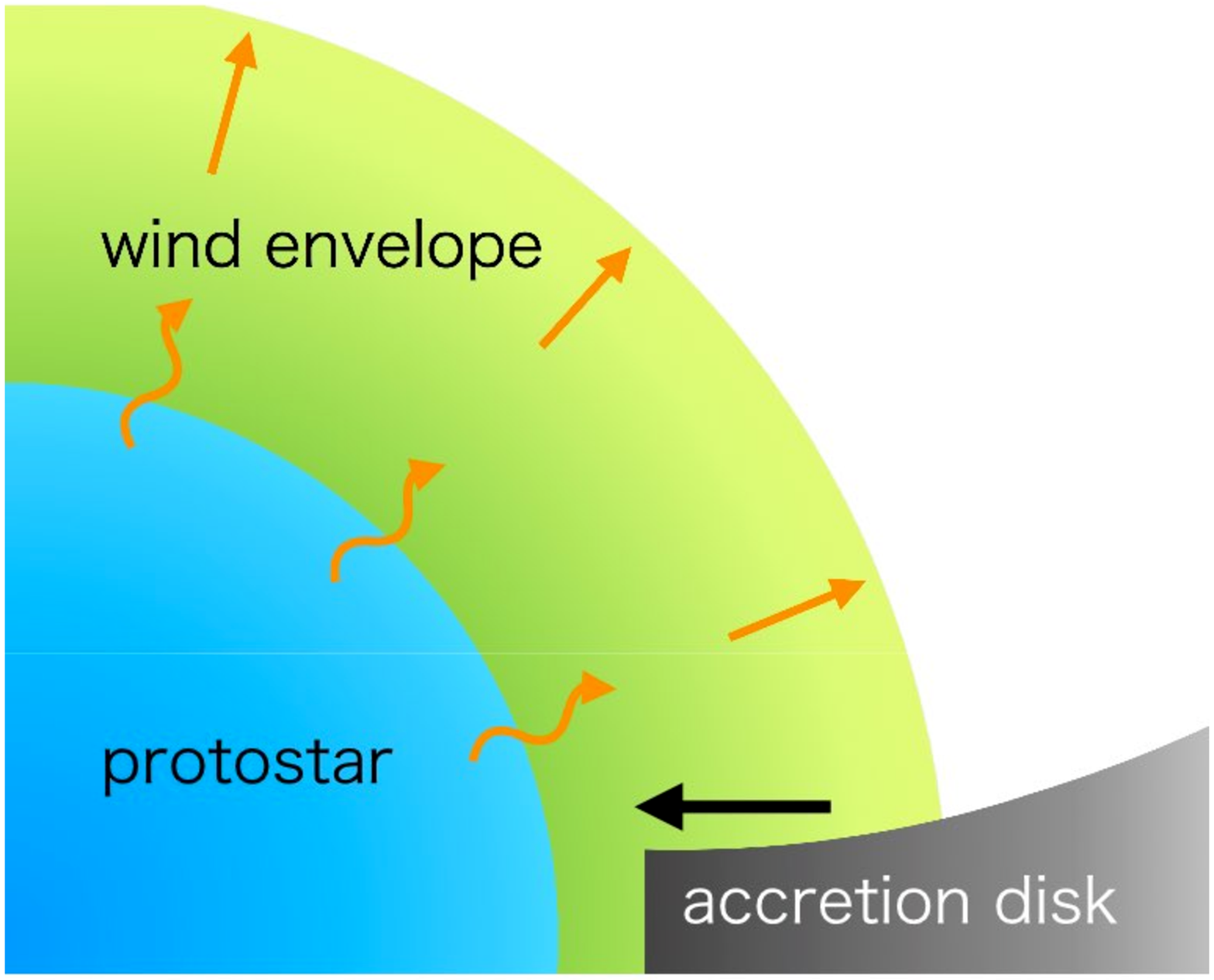}
\caption{Schematic picture of the stellar winds launched from the accreting SGPSs.}
\label{fig:schem_pic}
\end{center}
\end{figure}

In this section, we describe the basic equations and boundary conditions to calculate the stellar wind solutions from SGPSs.
In Figure \ref{fig:schem_pic}, we illustrate the situation considered here.
We suppose that the SGPS is composed of metal-free gas and gains the mass through the geometrically thin accretion disk.
Except for the equatorial region, stellar winds could be launched from the surface by the radiation pressure force.
We do not consider the interaction between the accretion disk and the wind, for simplicity.
Assuming that the accretion region is small in comparison with the outflowing region, 
we consider the steady wind structure under the assumption of the spherical symmetry.

\subsection{Basic Equations}\label{subsec:basic_eqs}

We focus on the formulation valid for the optically 
thick winds~\citep{Finzi1971,Zytkow1972,Quinn1985,Lee1990,Kato1992,Kato1994,Nugis2002,Dotan2012,Ro2016},
which is equivalent to assuming that the wind acceleration occurs beneath the photosphere
as in the case of the dense WR wind~\citep[e.g.,][]{Crowther2007}. 
The basic equations governing the wind structure are as follows. 

First, the equations of motion~(EoM) and continuity~(EoC) are
\begin{equation}
v \frac{dv}{dr} + \frac{1}{\rho} \frac{dP}{dr} + \frac{GM_\ast}{r^2} = 0,
\label{eq:eom}
\end{equation}
\begin{equation}
\dot{M}_{\rm wind} \equiv 4 \pi r^2 \rho v = {\rm const.},
\label{eq:eoc}
\end{equation}
where $G$ is the gravitational constant, $M_\ast$ the total stellar mass, $\dot{M}_{\rm wind}$ the mass outflow rate, and $v, P$, and $\rho$ correspond to the velocity, total pressure, and density at radius $r$, respectively.
Since the radiation field is the thermal black body in good approximation, 
the radiation pressure is given by $P_{\rm rad} = a T^4/3$, where $a$ is the radiation constant, and $T$ the temperature.
The total pressure $P$ is given by the sum of the gas pressure $P_{\rm gas}$ and the radiation pressure $P_{\rm rad}$:
\begin{equation}
P = P_{\rm gas} + P_{\rm rad} = \frac{\mathcal{R}}{\mu} \rho T + \frac{1}{3} a T^4,
\label{eq:eos}
\end{equation}
where $\mathcal{R}$ is the gas constant, and $\mu = \mu(\rho, T)$ the mean molecular weight.
We consider metal-free gas, composed of H and He with the mass fractions of
$X = 0.7$ and $Y = 0.3$, respectively.
The mean molecular weight $\mu$ varies with the ionization degrees of H, He, and He$^+$ in the wind.
After integration with respect to $r$, the energy equation is
\begin{equation}
\Lambda \equiv L_{\rm rad} + \dot{M}_{\rm wind} \left(\frac{v^2}{2} + w_{\rm gas} + w_{\rm rad} - \frac{GM_\ast}{r}\right) = {\rm const.},
\label{eq:energy}
\end{equation}
where $w_{\rm gas} = e_{\rm gas} + P_{\rm gas}/\rho$ is the specific enthalpy of the gas, with $e_{\rm gas}$ the specific internal energy of the gas including the ionization energy of H, He, and He$^+$, and $w_{\rm rad} = 4 a T^4 / 3 \rho$ that of radiation, respectively.
The radiative luminosity in the fluid frame $L_{\rm rad}$ is calculated by the diffusion approximation:
\begin{equation}
L_{\rm rad} = - \frac{16 \pi a c r^2 T^3}{3 \kappa \rho}\frac{dT}{dr},
\label{eq:l_rad}
\end{equation}
where $c$ is the speed of light, and $\kappa$ the Rosseland mean opacity.
For the Rosseland mean opacity, we use the tabulated values from 
the OPAL project~\citep{Iglesias1996} and from \citet{Alexander1994} above and below $7000\ {\rm K}$, respectively.

In Eqs. (\ref{eq:eom}-\ref{eq:l_rad}), we have four unknown functions, $v(r), \rho(r), T(r)$, and $L_{\rm rad}(r)$.
A wind solution can be obtained with the proper boundary conditions provided, which we describe in the next subsection.

\subsection{Boundary Conditions}\label{subsec:bc}
We impose the boundary conditions at the sonic point~$r_{\rm s}$ and at the matching point $r_{\rm m}$ of the star 
and the wind.
The latter corresponds to the base of the wind.

\subsubsection{Condition at the Sonic Point}

The sonic point corresponds to the singular point of the EoM~(Eq. \ref{eq:eom}).
Eq. \eqref{eq:eom} can be rewritten in a form that explicitly shows the presence of 
the singular point, by substituting the EoC~(Eq. \ref{eq:eoc}) and the EoS~(Eq. \ref{eq:eos}) into Eq. \eqref{eq:eom}:
\begin{equation}
\frac{1}{v} \frac{d v}{d r} = \left[\frac{2}{r}c_{\rm T}^2 - \frac{1}{\rho} \left(\frac{\partial P_{\rm gas}}{\partial T}\right)_\rho \frac{d T}{dr} + \frac{GM_\ast}{r^2}(\Gamma_{\rm r}-1)\right] / \left(v^2 - c_{\rm T}^2 \right),
\label{eq:vel_grad}
\end{equation}
where $c_{\rm T} = \sqrt{(\partial P/\partial \rho)_T}$ is the isothermal sound speed, 
and $\Gamma_{\rm r} \equiv L_{\rm rad}/L_{\rm Edd}$ the Eddington ratio.

In the wind solutions, the numerator of Eq. \eqref{eq:vel_grad} should vanish simultaneously at the sonic point, 
since the velocity gradient is required to be finite there~\citep{Lamers1999}.
This gives us the following boundary condition:
\begin{equation}
\Gamma_{\rm r} = \frac{1-\left(\frac{2 c_{\rm T}}{v_{\rm esc}}\right)^2}{1+\left(\frac{\partial P_{\rm gas}}{\partial P_{\rm rad}}\right)_\rho}\ \text{at}\ v = c_{\rm T},
\label{eq:cr_pt}
\end{equation}
where $v_{\rm esc} \equiv \sqrt{2 G M_\ast / r}$ is the escape velocity at radius $r$.

We obtain a unique stellar wind solution for each set of radius, density, and temperature 
$(r_{\rm s}, \rho_{\rm s}, T_{\rm s})$ at the sonic point.
The velocity $v_{\rm s}$ and the radiation luminosity $L_{\rm rad, s}$ at the sonic point 
are evaluated by using the first boundary condition~(Eq. \ref{eq:cr_pt}) as 
$v_{\rm s} = c_{\rm T}(\rho_{\rm s}, T_{\rm s})$ and 
$L_{\rm rad, s} = L_{\rm rad}(r_{\rm s}, \rho_{\rm s}, T_{\rm s})$, respectively.
Substituting the evaluated values of $v_{\rm s}$ and $L_{\rm rad, s}$ into Eqs. \eqref{eq:eoc} and \eqref{eq:energy},
we can fix $\dot{M}_{\rm wind}$ and $\Lambda$, which remain constant throughout a wind solution.
The velocity gradient at the sonic point is obtained by applying the de l'Hopital 
rule to Eq. \eqref{eq:vel_grad}~\citep{Lamers1999, Nugis2002}.
The above procedure allows us to construct one wind solution that smoothly passes through the sonic point.

\subsubsection{Conditions at the Matching Point of Star and Wind}

We impose another boundary condition at the base of the stellar wind to connect physical 
quantities continuously from the star to the wind.
We assume that the wind starts blowing at some radius $r_{\rm m}$.
We regard $r_{\rm m}$ as a free parameter without 
specifying how the wind initially arises in the atmosphere.
For a hydrostatic stellar model, we set the matching point from a layer 
that satisfies the following two conditions.
First, we require that the mass contained between the matching and the sonic radii is small and 
less than 5\ \% of the stellar mass.
This is because the mass is taken as a constant and equal to the stellar one in the gravity term of the wind equation.
Second, we require that, around the matching radius, the energy generation either via nuclear burning 
or gravitational contraction is negligible and the total luminosity becomes 
constant in radius~($L_{\rm r} \sim$ const.).

For the boundary conditions, we first require that the density and temperature are continuous 
across the matching point $r_{\rm m}$:
\begin{equation}
\rho(r_{\rm m}) = \rho_{\ast} (r_{\rm m}) \ \text{and}\ T(r_{\rm m}) = T_{\ast} (r_{\rm m}).
\label{eq:bc_rho_temp}
\end{equation}
where the subscript $_{\ast}$ indicates the quantities from the (hydrostatic) stellar model.
Note that the wind velocity is highly subsonic at $r_{\rm m}$ as long as $r_{\rm m} \ll r_{\rm s}$.
The first term in Eq. \eqref{eq:eom} thus being much smaller than the second term, 
i.e., $v |dv/dr| \ll \rho^{-1} |dP_{\rm gas}/dr|$, so that the gas is almost in the hydrostatic equilibrium around $r_{\rm m}$, i.e., 
the density and temperature of the wind asymptotically approach those of the hydrostatic stellar model.
Second, without the energy source in the envelope, the energy flux must be continuous 
across $r_{\rm m}$:
\begin{eqnarray}
L_{\ast}(r_{\rm m}) &=& \left[ L_{\rm rad} + \dot{M}_{\rm wind}\left(\frac{v^2}{2} + w_{\rm gas} + w_{\rm rad}\right)\right]_{r_{\rm m}}  \notag \\
&=& \Lambda + \frac{GM_\ast}{r_{\rm m}}\dot{M}_{\rm wind},
\label{eq:bc_energy_flux}
\end{eqnarray}
where $L_{\ast}(r_{\rm m})$ is the total luminosity of the star at $r_{\rm m}$.
Eq. \eqref{eq:bc_energy_flux} indicates that at the matching radius, $r_{\rm m}$, a fraction of the luminosity in the hydrostatic envelope, $L_{\ast}(r_{\rm m})$, is converted into the wind kinetic energy and the internal energy advected with the bulk motion of the flow. This leads to discontinuities in the gradients of temperature and density as well as in the radiation luminosity at $r_{\rm m}$.

\subsubsection{Constructing the Wind Solution Connected to a Stellar Model at a Matching Point}

With our four boundary conditions, one at the sonic point and three at the matching point, 
we can find four unknown functions $v(r), \rho(r), T(r)$, and $L_{\rm rad}(r)$ to 
construct a stellar wind solution continuously connected to the hydrostatic model.

The numerical integration is performed in the following way.
For a given matching point $r_{\rm m}$, we obtain the total luminosity $L_{\ast}(r_{\rm m})$ 
from the hydrostatic stellar model.
We guess the density and temperature at the sonic point ($\rho_{\rm s}$, $T_{\rm s}$) 
and find the sonic radius $r_{\rm s}$ by using the boundary condition Eq. \eqref{eq:bc_energy_flux} 
with $L_{\ast}(r_{\rm m})$.
Then, we integrate Eqs. (\ref{eq:eom}-\ref{eq:l_rad}) numerically inward from $r_{\rm s}$ to $r_{\rm m}$.
This is repeated with improving the guess for ($\rho_{\rm s}$, $T_{\rm s}$) until the two boundary 
conditions at $r_{\rm m}$~(Eq. \ref{eq:bc_rho_temp}) are satisfied.
At this moment, we obtain a unique wind solution in the subsonic region, as well as 
the constants of the motion, $\dot{M}_{\rm wind}$ and $\Lambda$.
The structure in the outer supersonic region can be solved as an initial value problem from 
the sonic point determined in the procedure above.

Eqs. (\ref{eq:eom}-\ref{eq:l_rad}) are solved by the explicit first-order Euler method.
The grid spacing is calculated from $\Delta r = \epsilon \times {\rm min} \left(T/ |dT/dr|, v/ |dv/dr|\right)$. We adopt $\epsilon = 10^{-4}$ as a fiducial value. We have confirmed that our results do not change with further reducing $\epsilon$.

\section{Classification of Wind Solutions}\label{sec:wind_general}

In this section, before discussing the proper wind solutions connected to the hydrostatic stellar model 
at the matching point, we see the general features of wind solutions. 
For this purpose, we here calculate a wind solution passing through the sonic point for each given set of
($r_{\rm s}, \rho_{\rm s}, T_{\rm s}$), and do not try to make it connect to the 
hydrostatic solution (see Section~\ref{sec:wind_formulation}). 

\begin{figure}
\begin{center}
\includegraphics[width=\columnwidth]{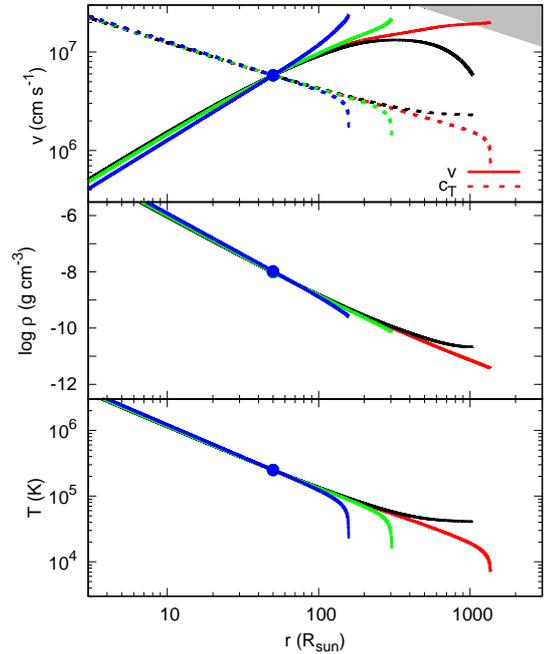}
\caption{Stellar wind solutions for $M_\ast = 100\ {\rm M}_\odot$, $T_{\rm s} = 2.5 \times 10^5\ {\rm K}$, and $r_{\rm s} = 50\ {\rm R}_\odot$.
In each panel, black, red, green, and blue lines correspond to $\rho_{\rm s} = 9.65, 9.71, 9.85$, and $10.5 \times 10^{-9}\ {\rm g}\ {\rm cm}^{-3}$, respectively.
The velocity (upper), density (middle) and temperature profiles (lower) are shown. 
The filled circle in each panel indicates the sonic point.
In the upper panel, the isothermal sound speed $c_{\rm T}$ (dashed) is also shown and 
the grey-shaded region indicates velocities exceeding the local escape velocity $v_{\rm esc}$.
}
\label{fig:wind_m100_t25}
\end{center}
\end{figure}

In Figure \ref{fig:wind_m100_t25}, we illustrate the stellar wind solutions for 
$M_\ast = 100\ {\rm M}_\odot$, $T_{\rm s} = 2.5 \times 10^5\ {\rm K}$, $r_{\rm s} = 50\ {\rm R}_\odot$ \footnote{We here adopt a very small sonic radius $r_{\rm s} \sim 50\ {\rm R}_\odot$ to compare the properties of the successful wind solution with those of the stalled one. If we adopt as large sonic radius~($r_{\rm s} \sim 2500\ {\rm R}_\odot$) as in the next section, we find that only stalled solutions are obtained.}, and four different values of density $\rho_{\rm s} = 9.65, 9.71, 9.85, 10.5 \times 10^{-9}\ {\rm g}\ {\rm cm}^{-3}$.
The velocity (top), density (middle), and temperature (bottom) profiles are shown.
For the different values of $\rho_{\rm s}$, the structure in the subsonic region
is similar to each other. 
On the other hand, the structure in the supersonic region largely differs, which
allows us to classify the solutions into the following two types:

i) {\it stalled wind solution}:

In some solutions, for example that with $\rho_{\rm s} = 9.65 \times 10^{-9}\ {\rm g}\ {\rm cm}^{-3}$, 
the velocity reaches the maximum at some radius and then decreases monotonically.
We call this type of solutions as the stalled wind solutions.
In this case, we stop the integration when the Mach number falls below 1.5.

ii) {\it ever accelerating wind solution}:

In some solutions, for example other cases ($\rho_{\rm s} \geq 9.71\times 10^{-9}\ {\rm g}\ {\rm cm}^{-3}$) 
shown in Figure \ref{fig:wind_m100_t25}, the velocity continues increasing monotonically 
up to $\sim 200\ {\rm km}\ {\rm s}^{-1}$.
We call this type of solutions as the ever accelerating wind solution.
In this case, we stop the integration at the photosphere, which is defined as the radius 
where the temperature becomes equal to the effective temperature 
$T_{\rm eff} = (L_{\rm rad}/4 \pi r^2 \sigma_{\rm SB})^{1/4}$, since our formalism is 
only valid in the optically thick regime, i.e., inside the photosphere.

\begin{figure}
\begin{center}
\includegraphics[width=\columnwidth]{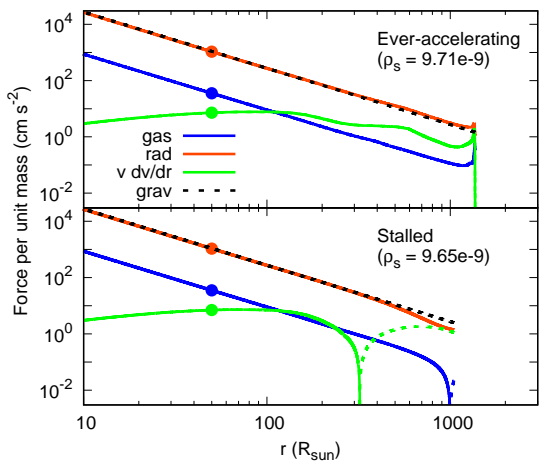}
\caption{Force balance in the wind region: the radiation pressure gradient~(red), 
gas pressure gradient~(blue), acceleration term~(green), and gravity~(black).
Shown is the case for $M_\ast = 100\ {\rm M}_\odot$, $T_{\rm s} = 2.5 \times 10^5\ {\rm K}$, and $r_{\rm s} = 50\ {\rm R}_\odot$. 
Depending on the density at the sonic point $\rho_{\rm s}$, the type of the wind solution changes: ({\it upper}; ever-accelerating solution) 
the case of $\rho_{\rm s} = 9.71 \times 10^{-9}\ {\rm g}\ {\rm cm}^{-3}$, ({\it lower}; stalled solution) $\rho_{\rm s} = 9.65 \times 10^{-9}\ {\rm g}\ {\rm cm}^{-3}$.
Dashed lines show that the force is in the negative (inward) direction.}
\label{fig:wind_m100_t25_acc}
\end{center}
\end{figure}

Figure \ref{fig:wind_m100_t25_acc} shows the force balance for the two types of the solutions 
$\rho_{\rm s} = 9.65 \times 10^{-9}\ {\rm g}\ {\rm cm}^{-3}$~(lower panel; stalled solution) 
and $9.71 \times 10^{-9}\ {\rm g}\ {\rm cm}^{-3}$~(upper panel; ever accelerating solution).
In the subsonic region, the acceleration term is much smaller than the total (gas and radiation) 
pressure gradient, i.e., the hydrostatic equilibrium holds well, in both cases.
Using the equation of hydrostatic equilibrium
\begin{equation}
\frac{1}{\rho} \frac{dP_{\rm gas}}{dr} = \frac{GM_\ast}{r^2}(\Gamma_{\rm r} - 1), 
\label{eq:dp_g/dr_stat}
\end{equation}
and the fact that the Eddington ratio $\Gamma_{\rm r}$ is almost constant 
because of the constancy of the opacity and radiative luminosity there, 
we can derive the power-law distributions $\rho \propto r^{-3}$ and $T \propto r^{-1}$ seen in Figure \ref{fig:wind_m100_t25}.
We also obtain $v \propto r$ from these relations and Eq. \eqref{eq:eoc}.

The wind structure in the supersonic region differs greatly among the stalled and ever-accelerating solutions.
In the stalled solution (Figure \ref{fig:wind_m100_t25_acc} lower), 
the outward pressure gradient has fallen below the inward pull of gravity already at $\sim 300\ {\rm R}_\odot$, 
so that the wind decelerates monotonically beyond this point. 
On the other hand, in the ever accelerating solution (Figure \ref{fig:wind_m100_t25_acc} upper), 
the acceleration continues until $\gtrsim 1000\ {\rm R}_\odot$ and reaches the velocity 
$\sim 200\ {\rm km}\ {\rm s}^{-1}$ at the photosphere, exceeding the escape velocity.
In this case, the wind is successfully launched with the mass-loss rate of 
$\dot{M}_{\rm wind} \sim 0.14\ {\rm M}_\odot\ {\rm yr}^{-1}$.

It should be noted, however, that not all the ever-accelerating solutions can be regarded as ``successful'' winds.
For example, for $\rho_{\rm s} = 9.85$ and $10.5 \times 10^{-9}\ {\rm g}\ {\rm cm}^{-3}$, 
although being continuously accelerated, the solution reaches the photosphere 
before the velocity exceeds the escape value.
Outside the photosphere, the further acceleration is improbable for the gas with the primordial composition~(see the discussion), 
and the matter in the wind would eventually fall back to the star.
Hence, the failure of some ever-accelerating solutions to launch the wind comes from the limited 
acceleration regime due to the small photospheric radii.

The successful wind solution has the most extended photosphere among the ever-accelerating solutions.
The maximum value for the photospheric radius can be estimated from the argument 
that the luminosity does not exceed the classical Eddington limit $L_{\rm Edd, es}$ 
and the effective temperature does not fall below $T_{\rm eff} \sim 5000\ {\rm K}$ due to the sharp opacity cutoff at lower temperatures:
\begin{eqnarray}
\label{eq:r_ph_max}
r_{\rm ph} & = & (L_{\rm ph}/4 \pi \sigma_{\rm SB} T_{\rm eff}^4)^{1/2} \\
&\lesssim& (L_{\rm Edd, es}/4 \pi \sigma_{\rm SB} T_{\rm eff}^4)^{1/2} \notag \\ 
& \lesssim & r_{\rm ph, max} \equiv 2160\ {\rm R}_\odot\ (M_\ast/100\ {\rm M}_\odot)^{1/2}\ (T_{\rm eff}/5000\ {\rm K})^{-2}. \notag
\end{eqnarray}
This value is consistent with our numerical result for 
$\rho_{\rm s} = 9.71 \times 10^{-9}\ {\rm g}\ {\rm cm}^{-3}$ with $r_{\rm ph} \sim 1400\ {\rm R}_\odot$~(red line, 
Figure \ref{fig:wind_m100_t25} middle).

We find that there is an upper limit on the stellar mass for the successful wind solutions to be found.
In these solutions, the velocity at the photosphere $v_{\rm ph}$ 
must be larger than the escape velocity there:
\begin{equation}
v_{\rm ph} \geq v_{\rm esc}(r_{\rm ph}).
\label{eq:wind_cond}
\end{equation}
From experiments, we found that the velocity at the photosphere $v_{\rm ph}$ is almost solely determined by 
$T_{\rm s}$~($v_{\rm ph} \sim 200\ {\rm km}\ {\rm s}^{-1}$ for our choice of $T_{\rm s} = 2.5 \times 10^5\ {\rm K}$ here) and increases with $T_{\rm s}$. 
Since they correspond to the solutions of the maximum photospheric radius,
by substituting Eq. \eqref{eq:r_ph_max} into Eq. \eqref{eq:wind_cond}, 
we obtain the upper limit on the mass of a star that can successfully launch the wind for a given $T_{\rm s}$, 
i.e., $v_{\rm ph}$:  
\begin{equation}
M_\ast \simeq 200\ {\rm M}_\odot\ (v_{\rm ph}/200\ {\rm km}\ {\rm s}^{-1})^4\ (T_{\rm eff}/5000\ {\rm K})^{-4}. 
\label{eq:M_star_max}
\end{equation}
For example, for $T_{\rm s} = 2.5 \times 10^5\ {\rm K}$ ($v_{\rm ph} \sim 200\ {\rm km}\ {\rm s}^{-1}$), 
the successful wind solutions exist only up to $M_\ast \simeq 200\ {\rm M}_\odot$.

\section{Wind Solutions Connected with the Hydrostatic Stars}\label{sec:wind_match}

In this section, we see whether SGPSs have the wind solutions with 
the terminal velocity exceeding the escape value.
We construct the solutions as described in Section \ref{subsec:bc}, i.e., 
by connecting the outer wind and inner hydrostatic solutions at the matching point.
As for the hydrostatic solutions, we adopt our previous results
of \citet{Hosokawa2013}, who followed the protostellar evolution 
until the stellar mass reaches $10^4\mbox{-}10^5\ {\rm M}_\odot$ at 
the constant accretion rates in the range 
$\dot{M}_{\rm acc} = 0.1\mbox{-}1.0\ {\rm M}_{\odot}\ {\rm yr}^{-1}$.
They showed that such protostars, whose structures are calculated 
under the assumption of hydrostatic equilibrium, have very extended 
envelopes. 
Below, we show that an outer part of the envelope can also take 
the outflowing structure, where the gas is not in the hydrostatic balance.

We first see the wind solutions for SGPSs accreting at $\dot{M}_{\rm acc} = 0.1\ {\rm M}_{\odot}\ {\rm yr}^{-1}$ in Section 4.1 
and then the cases with the higher accretion rate $\dot{M}_{\rm acc} = 1.0\ {\rm M}_{\odot}\ {\rm yr}^{-1}$ in Section 4.2. 
In the former case, the stellar models are available only up to the mass reaches $M_* \sim 10^4\ {\rm M}_\odot$, 
while in the latter case, up to $M_* \sim 10^5\ {\rm M}_\odot$.
Hence, for SGPSs more massive than $\sim 10^4\ {\rm M}_\odot$, wind solutions can be examined only in the latter case.

\subsection{Cases with $\dot{M}_{\rm acc} = 0.1\ {\rm M}_{\odot}\ {\rm yr}^{-1}$}

\subsubsection{$M_\ast = 100\ {\rm M}_\odot$ SGPS}\label{sssec:100Msun}
\citet{Hosokawa2013} showed that, at this accretion rate, 
by the time the stellar mass reaches $100\ {\rm M}_\odot$
the protostar already has the extended envelope of $R_\ast \simeq 2000\ {\rm R}_\odot$, 
characteristic to the SGPSs.
If the mass loss is vigorous enough to prevent the stellar growth at this early stage of the SGPSs, 
the star cannot reach the supermassive regime $M_* \sim 10^4\mbox{-}10^5~{\rm M}_\odot$.
We thus study the case of $M_\ast = 100\ {\rm M}_\odot$,
before considering the more massive regime in Section 4.1.3. 
Here, the matching point $r_{\rm m}$ is taken outside 
$1300~{\rm R}_\odot$, which encompasses more than 95\ \% of the total mass.

\begin{figure}
\begin{center}
\includegraphics[width=\columnwidth]{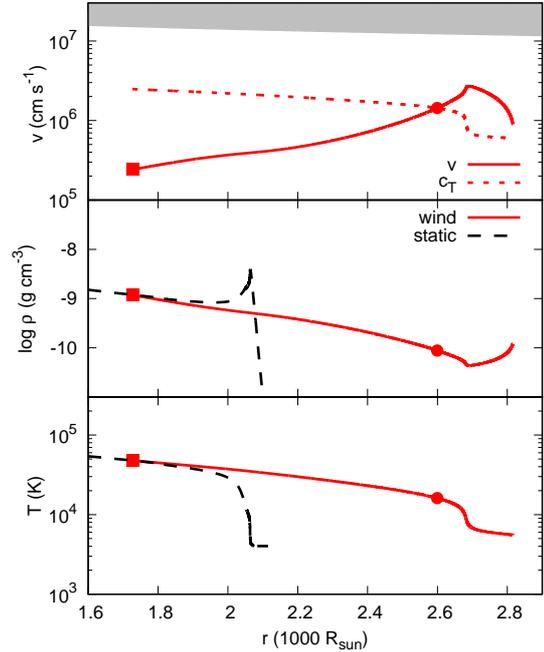}
\caption{The wind solution for $\dot{M}_{\rm acc} = 0.1\ {\rm M}_{\odot}\ {\rm yr}^{-1}$ and $M_\ast = 100\ {\rm M}_\odot$, 
with the matching point at $r_{\rm m} = 1700\ {\rm R}_\odot$~(the filled square).
In the upper panel, the wind velocity~(solid) is shown along with the isothermal sound speed ~(dotted). 
The upper grey-shaded region indicates the region where velocity exceeds the local escape velocity. 
The middle and lower panels show the density and temperature profiles, respectively.
Also shown in these panels are the profiles for the hydrostatic star~(dashed).}
\label{fig:vel_100}
\end{center}
\end{figure}

In Figure~\ref{fig:vel_100}, we show the wind solution with
the matching point at $r_{\rm m} = 1700\ {\rm R}_\odot$ (the filled square).
The velocity, density and temperature profiles are presented in the top, middle and 
bottom panels, respectively.
The flow is initially subsonic with the Mach number $< 0.1$ around
the matching point and then becomes supersonic at $r_{\rm s} \simeq 2600\ {\rm R}_\odot$.
The flow, however, decelerates after taking the maximum velocity at $\simeq 2700\ {\rm R}_\odot$, i.e., 
it is the stalled solution.
The wind solution~(red solid) has the more extended structure with the lower density  
than in the hydrostatic model~(black dashed).
Note that, unlike the hydrostatic case having the density inversion 
around $r = 2000~{\rm R}_\odot$, 
the wind solution does not have such a structure and the flow 
is just accelerated by the radiation pressure.

\begin{figure}
\begin{center}
\includegraphics[width=\columnwidth]{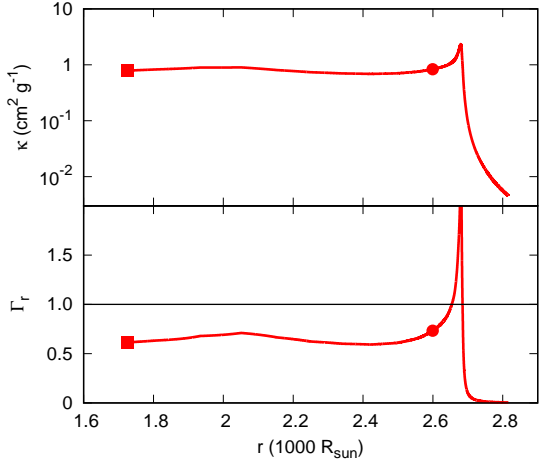}
\caption{Radial distributions of the Rosseland mean opacity $\kappa$~(upper) and the local Eddington ratio $\Gamma_{\rm r}$~(lower) 
for the wind solution with $\dot{M}_{\rm acc} = 0.1\ {\rm M}_{\odot}\ {\rm yr}^{-1}$ and $M_\ast = 100\ {\rm M}_\odot$.
The matching point (filled square) is located at $r_{\rm m} = 1700\ {\rm R}_\odot$. The sonic point is indicated by the filled circle.
Note that $\kappa$ is shown using a log scale, whereas $\Gamma_{\rm r}$ is shown using a linear scale.}
\label{fig:edd_100}
\end{center}
\end{figure}

Figure~\ref{fig:edd_100} shows the radial distributions 
of the Rosseland mean opacity $\kappa$~(upper panel) and 
the local Eddington ratio $\Gamma_{\rm r}$~(lower panel).
The radiative luminosity remains sub-Eddington, i.e., 
$\Gamma_{\rm r}<1$, throughout the subsonic region. 
Around the opacity bump at $\simeq 2700\ {\rm R}_\odot$ due to 
the bound-free absorption of H atoms and ${\rm H}^{-}$ ions, 
a thin super-Eddington layer appears just outside the sonic point. 
The sharp drop of the opacity caused by the hydrogen recombination 
below $10^4$~K, however, pushes back the Eddington ratio 
below unity again for $r \gtrsim 2700\ {\rm R}_\odot$.

\begin{figure}
\begin{center}
\includegraphics[width=\columnwidth]{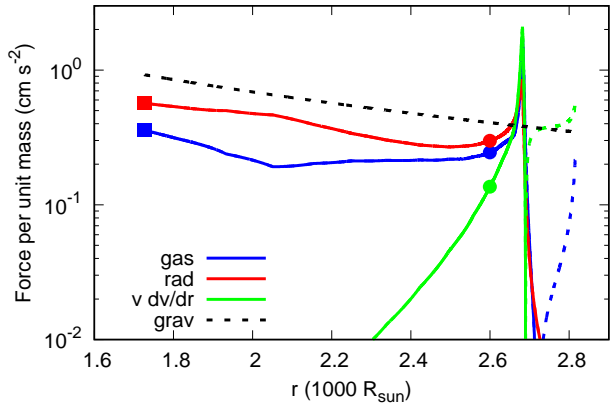}
\caption{Same as Figure~\ref{fig:wind_m100_t25_acc}, but for the wind solution matched with the hydrostatic envelope: 
the radiation pressure gradient~(red), gas pressure gradient~(blue), acceleration term~(green), and gravity~(black).
The dashed parts of the lines show that the force is in the negative (inward) direction.
The stellar parameters are $\dot{M}_{\rm acc} = 0.1\ {\rm M}_{\odot}\ {\rm yr}^{-1}$ and $M_\ast = 100\ {\rm M}_\odot$.
The matching point (filled square) is located at $r_{\rm m} = 1700\ {\rm R}_\odot$. The sonic point is indicated by the filled circle.}
\label{fig:force_100}
\end{center}
\end{figure}

Figure~\ref{fig:force_100} shows the force balance for this case.
We see that the radiation pressure~(red) dominates the gas pressure~(blue) everywhere.
Since the pressure gradient is almost in balance with
the gravity~(black), hydrostatic equilibrium still holds
approximately in the subsonic region.
The acceleration term~(green) gradually increases outward
and takes the maximum in the supersonic region
at $r \simeq 2700\ {\rm R}_\odot$.
However, the acceleration term then declines dramatically
in the outer region owing to the decrease of the radiation pressure force, 
which is in proportion to the opacity $\kappa$.

\subsubsection{Dependence on the Matching Radius}

\begin{figure}
\begin{center}
\includegraphics[width=\columnwidth]{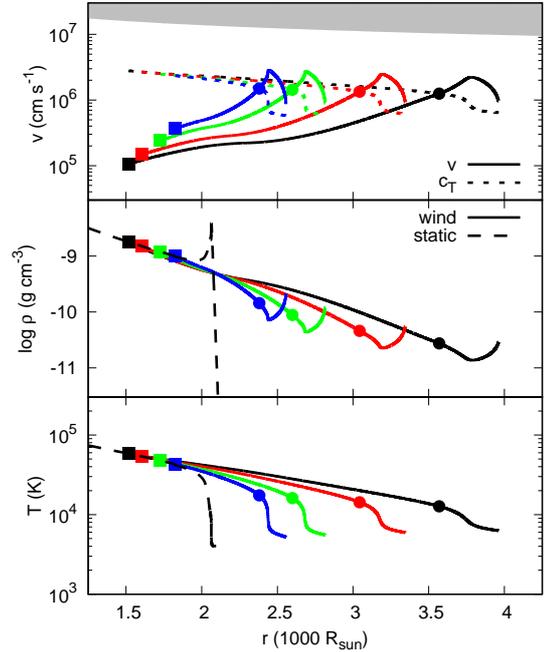}
\caption{Effect of the different matching radius on the wind structure. 
Shown are the velocity (upper), density (middle), and temperature (lower) profiles 
for the solutions with the same stellar parameters
$\dot{M}_{\rm acc} = 0.1\ {\rm M}_{\odot}\ {\rm yr}^{-1}$ and $M_\ast = 100\ {\rm M}_\odot$
but with the different matching radii $r_{\rm m} =$ 1800~(blue), 1700~(green), 1600~(red), 
and $1500\ {\rm R}_\odot$~(black).}
\label{fig:vel_100_dif}
\end{center}
\end{figure}

\begin{figure}
\begin{center}
\includegraphics[width=\columnwidth]{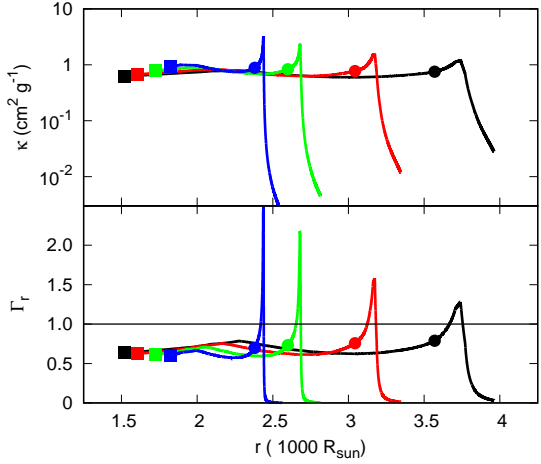}
\caption{Profiles of the Rosseland mean opacity $\kappa$~(upper) and the local Eddington ratio 
$\Gamma_{\rm r}$~(lower) for the wind solution with the same stellar parameters 
$\dot{M}_{\rm acc} = 0.1\ {\rm M}_{\odot}\ {\rm yr}^{-1}$ and $M_\ast = 100\ {\rm M}_\odot$ but with the 
different values of the matching radii $r_{\rm m} =$ 1800 (blue), 1700 (green), 1600 (red), and $1500\ {\rm R}_\odot$ (black). 
Note that $\kappa$ is shown using a log scale, whereas $\Gamma_{\rm r}$ is shown using a linear scale.}
\label{fig:Edd_100_dif}
\end{center}
\end{figure}

Next, we investigate how the wind structure changes
by varying the matching radius $r_{\rm m}$.
Figures \ref{fig:vel_100_dif} and \ref{fig:Edd_100_dif} present
the wind solutions for the different matching radii with the same
$M_* = 100~{\rm M}_\odot$ SGPS model.
In each figure, the blue, green, red, and black lines 
represent those with the different matching radii of
$r_{\rm m} = 1800, 1700, 1600$, and $1500\ {\rm R}_\odot$, respectively.

The upper panel of Figure \ref{fig:vel_100_dif} shows that
the velocity structure is qualitatively similar to each other even with the different matching radii: 
the flow velocity increases and becomes supersonic at some point, but starts decreasing before 
exceeding the escape value.
This is due to the sharp decrease of the opacity and radiation force caused by the recombination of 
hydrogen~(upper panel of Figure~\ref{fig:Edd_100_dif}), as mentioned in Section~\ref{sssec:100Msun}.

With the smaller $r_{\rm m}$, the flows have the mass-loss rates $\dot{M}_{\rm wind} = 1.2, 0.83, 0.55, 0.42\ {\rm M}_\odot\ {\rm yr}^{-1}$, which could have a great impact on the stellar growth if the wind is successfully launched.
All of them, however, belong to the stalled wind solutions, which implies that the steady wind 
is not launched from this stellar model regardless of the matching radius, and that 
the stellar mass acquisition via accretion is thus not prevented by the wind mass loss.

In the middle panel of Figure~\ref{fig:vel_100_dif}, we can see that,
with the smaller matching radius $r_{\rm m}$, the outflowing envelope 
has the more extended structure with the lower density at the sonic point.
On the other hand, the temperature at the sonic point, which is located just inside 
the opacity peak (see Figure~\ref{fig:Edd_100_dif} upper), is $\simeq 10^4\ {\rm K}$ 
for all the cases (Figure \ref{fig:vel_100_dif} lower) 
because of the very strong temperature-dependence of the opacity around this value.

The maximum value of the local Eddington ratio $\Gamma_{\rm r, max}$ 
is lower for the inner matching point case~(Figure~\ref{fig:Edd_100_dif} lower).
This is because the density above the sonic point $r > r_{\rm s}$ becomes lower for the smaller $r_{\rm m}$, 
which results in the lower opacity and thus the smaller $\Gamma_{\rm r, max}$. 
In particular, $\Gamma_{\rm r, max}$ never reaches unity for $r_{\rm m}$ smaller than $1500\ {\rm R}_\odot$. 

\subsubsection{Dependence on the Stellar Mass}

\begin{figure}
\begin{center}
\includegraphics[width=\columnwidth]{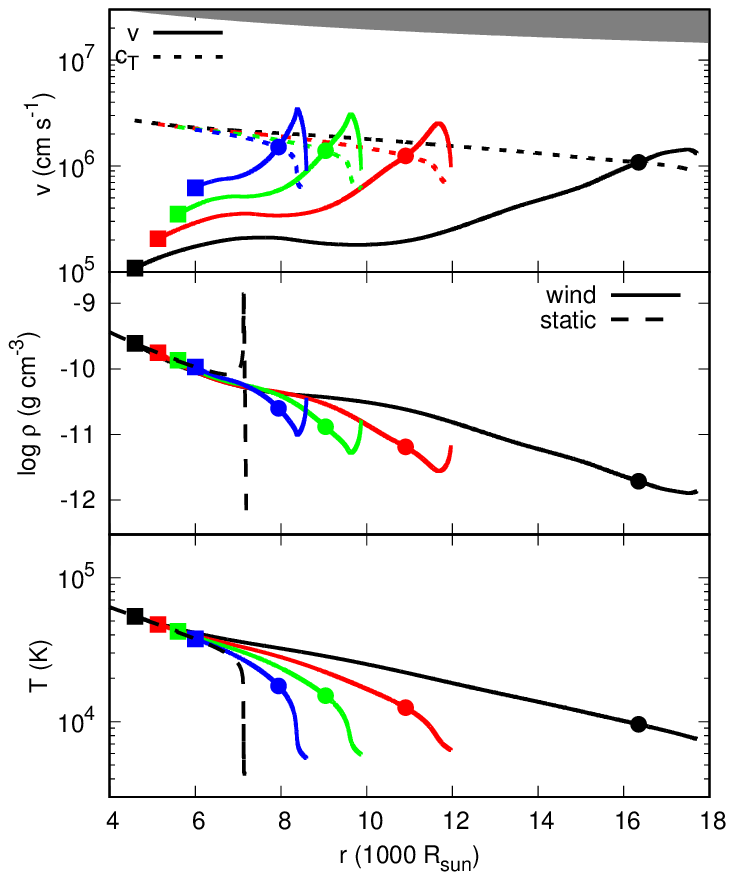}
\caption{Same as Figure \ref{fig:vel_100_dif}, but for the stellar model with $\dot{M}_{\rm acc} = 0.1\ {\rm M}_{\odot}\ {\rm yr}^{-1}$ and $M_\ast = 1000\ {\rm M}_\odot$.
In each panel, the blue, green, red, and black lines correspond to the results for $r_{\rm m} = 6000, 5500, 5000$, and $4500\ {\rm R}_\odot$, respectively.}
\label{fig:vel_01_1000_dif}
\end{center}
\end{figure}

\begin{figure}
\begin{center}
\includegraphics[width=\columnwidth]{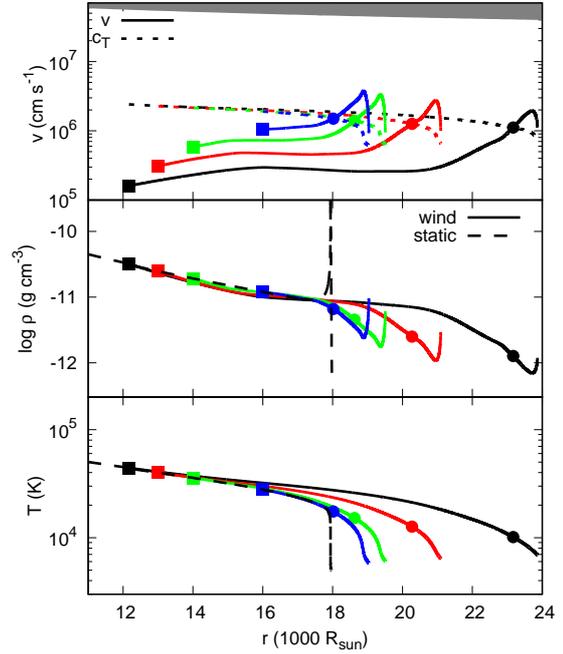}
\caption{Same as Figure \ref{fig:vel_100_dif}, but for the stellar model with $\dot{M}_{\rm acc} = 0.1\ {\rm M}_{\odot}\ {\rm yr}^{-1}$ and $M_\ast = 10^4\ {\rm M}_\odot$.
In each panel, the blue, green, red, and black lines correspond to the results for $r_{\rm m} = 16000, 14000, 13000$, and $12000\ {\rm R}_\odot$, respectively.}
\label{fig:vel_01_1e4_dif}
\end{center}
\end{figure}

Next, we examine the stellar wind solutions for more 
massive SGPSs with the same accretion rate $\dot{M}_{\rm acc} = 0.1\ {\rm M}_{\odot}\ {\rm yr}^{-1}$.
The wind solutions for the 1000 and $10^{4}\ {\rm M}_\odot$ stars 
are shown in Figures~\ref{fig:vel_01_1000_dif} and \ref{fig:vel_01_1e4_dif}, respectively.
Both stellar models have the extended envelopes with the radii
$R_\ast \simeq 7000\ {\rm R}_\odot$~(1000\ ${\rm M}_\odot$) and $18000\ {\rm R}_\odot$~($10^{4}\ {\rm M}_\odot$).
The 95\ \% of the total stellar mass is enclosed within $2000\ {\rm R}_\odot$ for the 1000~${\rm M}_\odot$ model 
and $500\ {\rm R}_\odot$ for the $10^{4}\ {\rm M}_\odot$ model, respectively.
The matching radii are chosen at $r_{\rm m} = 6000, 5500, 5000$, and $4500\ {\rm R}_\odot$ 
for 1000~${\rm M}_\odot$, and $r_{\rm m} = 16000, 14000, 13000$, and $12000\ {\rm R}_\odot$ for 
$10^{4}\ {\rm M}_\odot$, respectively.

All these solutions for the 1000 and $10^{4}\ {\rm M}_\odot$ stars are again the stalled ones
as in the case of the $100\ {\rm M}_\odot$ model: 
the flow starts to decelerate after reaching the sonic point without reaching the escape velocity. 
Note also that the maximum velocity in the wind remains much below the escape velocity $v_{\rm esc, \ast}$ 
for more massive models
since the escape velocity increases with the stellar mass as $v_{\rm esc, \ast} \propto M_\ast^{1/4}$
from the relation $R_\ast \propto M_\ast^{1/2}$ for the SGPSs \citep{Hosokawa2012b}.
Whereas the mass-loss rates are mathematically determined as $\dot{M}_{\rm wind} = 2.3, 1.4, 0.92, 0.54\ {\rm M}_\odot\ {\rm yr}^{-1}$~($\dot{M}_{\rm wind} = 3.1, 2.1, 1.3, 0.73\ {\rm M}_\odot\ {\rm yr}^{-1}$) for the 1000~${\rm M}_\odot$~($10^{4}\ {\rm M}_\odot$) models, such steady winds can not be launched from these stars and prevent the mass growth.

For the flow to be accelerated to the supersonic regime, the matching point must be located outside a certain radius, 
which is $r_{\rm m, min} \simeq 4500 \ {\rm R}_\odot~(12000\ {\rm R}_\odot)$ 
for the $M_\ast = 1000 \ {\rm M}_\odot~(10^4 \ {\rm M}_\odot)$ case. 
As seen in Figures~\ref{fig:vel_01_1000_dif} and \ref{fig:vel_01_1e4_dif}, 
the velocity gradient at the sonic point becomes smaller for the smaller matching radius 
and it eventually becomes even negative below the threshold value $r_{\rm m, min}$.
In this case, the flow cannot reach the supersonic regime, so that we here consider only the case of $r_{\rm m} > r_{\rm m, min}$. 

Hence, with the accretion rate of $\dot{M}_{\rm acc} = 0.1\ {\rm M}_{\odot}\ {\rm yr}^{-1}$, 
the mass loss by stellar winds does not prevent SGPSs from growing up at least to $10^4\ {\rm M}_\odot$.
Without more massive SGPS models, we can not examine the effect of stellar winds on the SGPS evolution for $M_* \geq 10^4\ {\rm M}_\odot$.
We expect, however, that a SGPS reaches the supermassive regime, since successful wind solutions 
exist only for $M_* \lesssim 200\ {\rm M}_\odot$, according to the analytical estimate in Section \ref{sec:wind_general}~(Eq. \ref{eq:M_star_max}).

\subsection{Cases with $\dot{M}_{\rm acc} = 1.0\ {\rm M}_{\odot}\ {\rm yr}^{-1}$}

\begin{figure}
\begin{center}
\includegraphics[width=\columnwidth]{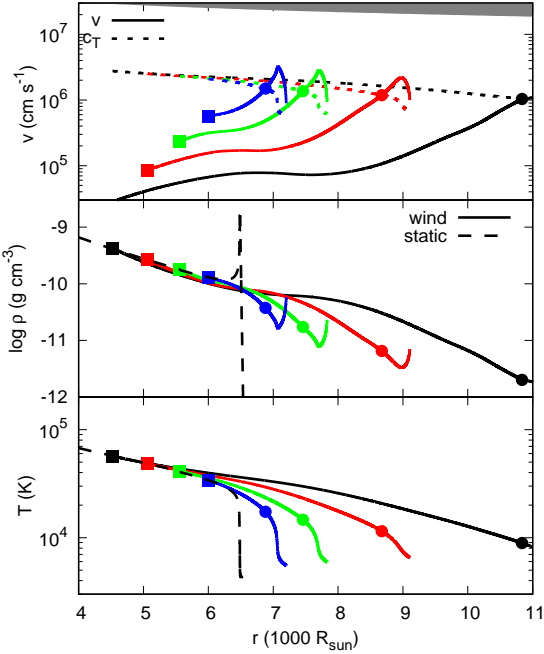}
\caption{Same as Figure \ref{fig:vel_100_dif}, but for the stellar model with $\dot{M}_{\rm acc} = 1.0\ {\rm M}_{\odot}\ {\rm yr}^{-1}$ and $M_\ast = 1000\ {\rm M}_\odot$.
In each panel, the blue, green, red, and black lines correspond to the results with $r_{\rm m} = 6000, 5500, 5000$, and $4500\ {\rm R}_\odot$, respectively.}
\label{fig:vel_1_1000_dif}
\end{center}
\end{figure}

\begin{figure}
\begin{center}
\includegraphics[width=\columnwidth]{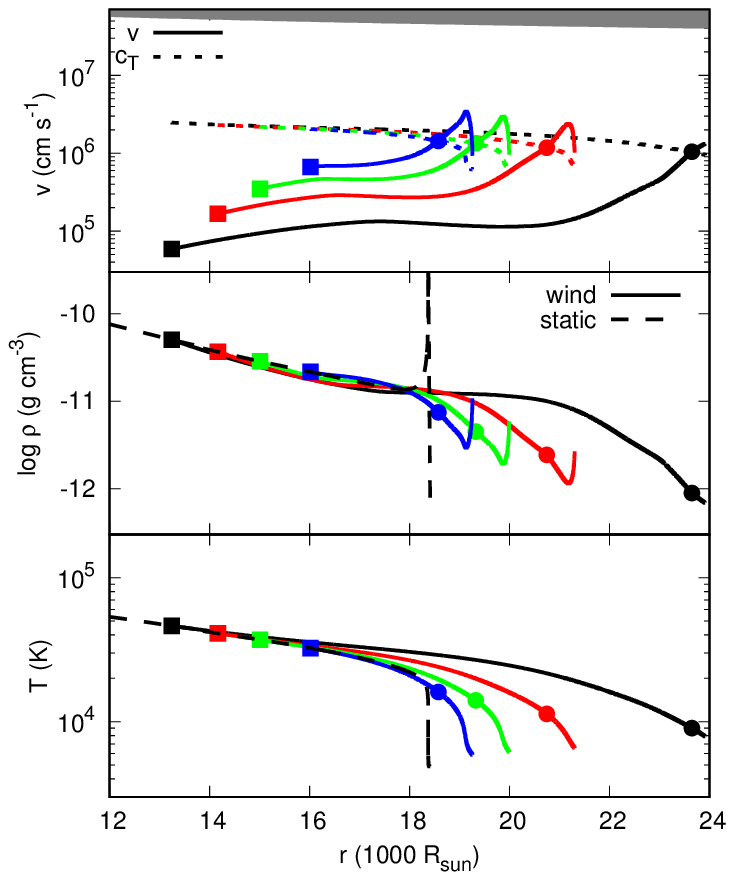}
\caption{Same as Figure \ref{fig:vel_100_dif}, but for the stellar model with $\dot{M}_{\rm acc} = 1.0\ {\rm M}_{\odot}\ {\rm yr}^{-1}$ and $M_\ast = 10^4\ {\rm M}_\odot$.
In each panel, the blue, green, red, and black lines correspond to the results with $r_{\rm m} = 16000, 15000, 14000$, and $13000\ {\rm R}_\odot$, respectively.}
\label{fig:vel_1_1e4_dif}
\end{center}
\end{figure}

\begin{figure}
\begin{center}
\includegraphics[width=\columnwidth]{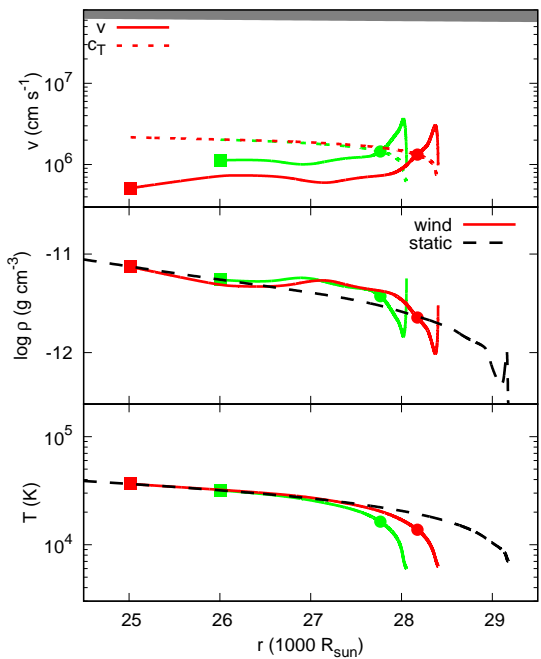}
\caption{Same as Figure \ref{fig:vel_100_dif}, but for the stellar model with $\dot{M}_{\rm acc} = 1.0\ {\rm M}_{\odot}\ {\rm yr}^{-1}$ and $M_\ast = 10^5\ {\rm M}_\odot$.
In each panel, the green and red lines correspond to the results with $r_{\rm m} = 26000$ and $25000\ {\rm R}_\odot$, respectively.
In this figure, the grey-shaded region show the region where $v \geq v_{\rm esc}/2$ holds.}
\label{fig:vel_1_1e5_dif}
\end{center}
\end{figure}

Here, we consider the cases with the higher accretion rate of $\dot{M}_{\rm acc} = 1.0\ {\rm M}_{\odot}\ {\rm yr}^{-1}$.
In this case, the protostar becomes a SGPS when the stellar mass exceeds $\gtrsim 100\ {\rm M}_\odot$~\citep{Hosokawa2013}.
Below, we show the results for the SGPS models with three different masses $M_\ast = 1000, 10^4$, and $10^5\ {\rm M}_\odot$.

Figures \ref{fig:vel_1_1000_dif} and \ref{fig:vel_1_1e4_dif} show the velocity, density, and temperature profiles 
of the wind solutions for the $M_\ast = 1000\ {\rm M}_\odot$ and $10^4\ {\rm M}_\odot$ models, respectively.
They have the extended envelopes with the radii
$R_\ast \simeq 6500\ {\rm R}_\odot$~(1000\ ${\rm M}_\odot$) and $18000\ {\rm R}_\odot$~($10^{4}\ {\rm M}_\odot$), and
$\sim$ 95\ \% of the total stellar mass is encompassed within $3500\ {\rm R}_\odot$~(1000\ ${\rm M}_\odot$) 
and $4000\ {\rm R}_\odot$~($10^{4}\ {\rm M}_\odot$), respectively.
The matching radii are chosen at four different radii: $r_{\rm m} = 6000, 5500, 5000$, and $4500\ {\rm R}_\odot$ 
for 1000~${\rm M}_\odot$ and $r_{\rm m} = 16000, 15000, 14000$, and $13000\ {\rm R}_\odot$ for $10^{4}\ {\rm M}_\odot$, respectively.
If we take the matching radii below $r_{\rm m, min} \sim 4500\ {\rm R}_\odot$~($\sim 13000\ {\rm R}_\odot$) for the 1000\ ${\rm M}_\odot$~($10^{4}\ {\rm M}_\odot$) model, 
the velocity gradient at the sonic point would finally be negative, and supersonic wind solutions can not be found,
as we discuss in Sections 4.1.2 and 4.1.3.

The wind solutions have both quantitatively and qualitatively similar structures 
as those with the lower accretion rate $\dot{M}_{\rm acc} = 0.1\ {\rm M}_{\odot}\ {\rm yr}^{-1}$
(c.f., Figures \ref{fig:vel_01_1000_dif} and \ref{fig:vel_01_1e4_dif}).
They have almost the same values for the mathematically determined mass-loss rates, as well:
$\dot{M}_{\rm wind} = 2.6, 1.3, 0.56, 0.24\ {\rm M}_\odot\ {\rm yr}^{-1}$~($\dot{M}_{\rm wind} = 3.6, 2.2, 1.2, 0.50\ {\rm M}_\odot\ {\rm yr}^{-1}$) for the 1000~${\rm M}_\odot$~($10^{4}\ {\rm M}_\odot$) model.
This is because the SGPS models at the same mass have the
similar envelope structures and radii regardless of the accretion rates.
They are all classified into the stalled wind solution, i.e., wind acceleration is stopped 
before reaching the escape velocity in the supersonic region, and the flow can not escape from the system steadily.
Hence, SGPSs can grow up to $10^4\ {\rm M}_\odot$ without being interrupted by stellar winds.

Finally in Figure \ref{fig:vel_1_1e5_dif}, we show the wind solutions which are obtained for the $M_\ast = 10^5\ {\rm M}_\odot$ model.
This SGPS model has a radius of $R_\ast \sim 29000\ {\rm R}_\odot$, and $\sim$ 95\ \% of the mass is enclosed
within $\sim 2500\ {\rm R}_\odot$.
The cases with the matching radii at $r_{\rm m} = 26000$ and $25000\ {\rm R}_\odot$ are shown.
Unlike all the wind solutions discussed so far, the radial extent of these wind solutions is shorter than the photospheric radius of the original
hydrostatic SGPS model.
They have the largest mass-loss rates: $\dot{M}_{\rm wind} = 4.0, 2.3\ {\rm M}_\odot\ {\rm yr}^{-1}$.
In these cases, again, the flows are classified into the stalled wind solution and fail to steadily escape from the gravitational pull of the star.
In conclusion, with the higher accretion rate of $\dot{M}_{\rm acc} = 1.0\ {\rm M}_{\odot}\ {\rm yr}^{-1}$, 
the mass loss by the stellar wind does not prevent 
the growth of a SGPS via mass accretion at least until a $10^5\ {\rm M}_\odot$ SMS forms.

\section{Summary and Discussion}\label{sec:summary}
We have examined whether in the supergiant protostar (SGPS) phase, a rapidly accreting protostar with the primordial composition has a steady 
optically thick wind that could cause such significant mass loss as to prevent the stellar mass growth.
We have constructed the steady wind solutions which are continuously connected to the 
hydrostatic stellar envelopes.
Our results show that the outflow stalls just after 
passing through the sonic point since the acceleration by radiation pressure becomes 
inefficient in the outermost part with temperature $\lesssim 10^4\ {\rm K}$, 
where the opacity sharply drops as hydrogen ions rapidly recombine.
The flow velocity does not reach the stellar escape velocity in any of the cases.
Hence, the growth of the SGPS mass via rapid accretion
will not be hindered by strong mass loss at least
until the SGPS mass reaches $10^4\mbox{-}10^5\ {\rm M}_\odot$.

Although the steady stellar wind is unlikely for the metal-free SGPSs, 
non-steady or sporadic mass loss might happen if we relax the assumption of the steady flow.
This situation is similar to the wind solutions exceeding the so-called photon-tiring limit that sets the maximum rate
of radiatively driven mass loss from hot massive stars~\citep{Owocki1997}.
In this case, a steady wind is not possible because the mass-loss rate is so high that the wind velocity never reaches the escape velocity
even if all the stellar luminosity is expended for wind acceleration, and the wind necessarily stagnates at some radius.
\citet{Marle2009} performed one-dimensional radiation hydrodynamical~(RHD) 
simulations to study the evolution of the photon-tired wind, and found 
that the sporadic outflow is in fact occasionally launched to escape 
from the stellar gravitational pull.
However, there is a notable difference in our and their winds.  
The opacity is always dominated by the Thomson scattering throughout their winds and remains at a 
constant value, while it drops sharply in the outermost part of the SGPS wind with $\lesssim 10^4$~K. 
Future time-dependent calculations are awaited to see
the effects of non-steady winds for the SGPS cases. 
 
The optically thick winds have originally been proposed 
to explain the substantial mass loss from 
WR stars~\citep[e.g.,][]{Kato1992, Nugis2002, Ro2016}.
Note that our results here are consistent with the previous studies of the WR winds,
although we have shown that the steady optically thick winds 
will be unimportant for the metal-free SGPSs.
For instance, \citet{Ro2016} constructed the optically thick wind solutions 
that are connected to the hydrostatic WR structure at their bases, 
in the same way as our SGPS winds above. 
They showed that the outflow is accelerated by radiation pressure, 
in particular, in the region with temperature 
$\simeq 2 \times 10^5\ {\rm K}$, where the opacity takes 
the large values owing to a number of bound-bound transitions in iron
(so-called the iron opacity ``bump'').
Nonetheless, none of their solutions achieves a high enough 
terminal velocity to exceed the stellar escape velocities, i.e., 
the outflow stalls as in our SGPS cases. 
They attributed this failure of reaching the escape velocity 
to their negligence of the contribution to opacity by the lines 
including the effect of velocity gradient, 
which enables more efficient momentum transport 
from radiation to matter~\citep[e.g.,][]{Castor1975}. 
In metal-free SGPS cases, however, this line force, which 
is proportional to the number of available lines, 
would be negligible because the number of lines is 
very limited for the pure hydrogen and helium composition
~\citep{Krticka2006}.

We have shown for the metal-free SGPSs that wind acceleration is rapidly quenched in the region of hydrogen recombination.
With the solar composition, however, it would be maintained even for cool massive main-sequence stars for which the hydrogen recombination occurs in their atmosphere. This is due to the radiation pressure exerted through spectral lines of iron. Observations show such winds have smaller terminal velocities and higher mass-loss rates than those from hotter massive stars~\citep[the so-called bistability jump; e.g.,][]{Lamers1995}. Therefore, this effect should be considered when we consider solar-metallicity wind models.

Multi-dimensional effects are known to be important 
in the surface layers of a massive star, 
where density inversion appears in the 1D hydrostatic model.
For the solar composition, density inversion could be developed by the iron opacity bump, and
the multi-dimensional structure of such layers was studied 
by \citet{Jiang2015} by way of the 3D local RHD simulations.
They showed that these layers are convectively unstable and
large density fluctuations give rise to a porous atmosphere, 
which is considered by some authors to play a key role in driving  
strong winds observed for WR stars and luminous 
blue variables~\citep[e.g.,][]{Shaviv2000,Marle2009}.
For SGPSs with the primordial composition, similar numerical simulations are needed, 
since density inversion is developed by the H and H$^-$ opacity, and it is uncertain whether 
it leads to a porous atmosphere which might promote wind driving.
Their calculation, however, was limited to the local patch of the stellar envelope, 
rather than encompassing the global wind structure, and so the resulting 
mass-loss rate was not predicted.
Therefore, global simulations are awaited to clarify the 
multi-dimensional effects on the wind driving in SGPS envelopes.

We have neglected any convective energy transport in the wind.
We expect that this will not largely modify our results from the following consideration.
In the supersonic regime, the convective energy flux is much less 
than the advected internal energy flux since the convection velocity 
is below the sound speed.
In the subsonic regime, if we include the convective contribution, 
the radiative luminosity would be reduced for the constant total luminosity 
and so as the radiative acceleration of a wind.
Therefore, the inclusion of convection should not change our conclusions that 
the steady wind is not driven from a SGPS.

\section*{Acknowledgments}
The authors thank K. Sugimura and T. Suda for fruitful discussions
and T. Sakurai for providing us with the hydrostatic stellar models.
This work is supported in part by the Grant-in-Aid from the Ministry of Education, Culture, Sports, Science and Technology (MEXT) of Japan, Nos.16J02951 (DN), 25800102, 15H00776 and 16H05996 (TH),
25287040 (KO), and 26400222 and 16H02168 (KN).

%%%%%%%%%%%%%%%%%%%%%%%%%%%%%%%%%%%%%%%%%%%%%%%%%%



\bibliographystyle{mnras}
%\bibliography{ref} % if your bibtex file is called example.bib


%\begin{thebibliography}{99}
%\end{thebibliography}

%%%%%%%%%%%%%%%%%%%%%%%%%%%%%%%%%%%%%%%%%%%%%%%%%%

%%%%%%%%%%%%%%%%% APPENDICES %%%%%%%%%%%%%%%%%%%%%

%\appendix



%%%%%%%%%%%%%%%%%%%%%%%%%%%%%%%%%%%%%%%%%%%%%%%%%%

% Don't change these lines
\bsp	% typesetting comment
\label{lastpage}
\end{document}